\documentclass[preprint,aps,showpacs,nofootinbib,superscriptaddress,preprintnumbers,epsf,psf]{revtex4}
\usepackage[english]{babel}
\usepackage{pstricks}
\usepackage[dvips]{graphicx}
\usepackage{epsf}
\usepackage{amsmath}
\usepackage{amsfonts}
\usepackage{amssymb}

\usepackage{color}
\usepackage{dcolumn}% Align table columns on decimal point
\usepackage{bm}% bold math
\everymath{\displaystyle}
\begin{document}
%%%%%%%%%%%%%%%%%%%%%%%%%%%%%%%%%%%%%%%%%%%%%%%%%%%%%%%%%%%%%%%%%%%%
% End of preamble and beginning of text.
%\pagestyle{empty}
%%%%%%%%%%%%%%%%%%%%%%%%%%%%%%%%%%%%%%%%%%%%%%%%%%%%%%%%%%%%%

\title{Axial anomaly and energy dependence of hyperon polarization in Heavy-Ion Collisions }

\author{\firstname{Alexander}~\surname{Sorin}}
\email{sorin@theor.jinr.ru}
\author{\firstname{Oleg}~\surname{Teryaev}}
\email{teryaev@theor.jinr.ru} \affiliation{Joint Institute for
Nuclear Research, 141980 Dubna (Moscow region), Russia}
\affiliation{National Research Nuclear University MEPhI (Moscow
Engineering Physics Institute), Kashirskoe Shosse 31, 115409 Moscow,
Russia} \affiliation{Dubna International University, 141980, Dubna,
Russia}
\date{\today}

\begin {abstract}
We address the issue of energy and charge dependence of global
polarization of $\Lambda$ hyperons in peripheral $Au-Au$ collisions
recently observed by the STAR collaboration at RHIC. We compare
different contributions to the anomalous mechanism relating
polarization to vorticity and hydrodynamic helicity in QCD matter.
We stress that the suppression of the gravitational anomaly
contribution in strongly correlated matter observed in lattice
simulations confirms our earlier prediction of rapid decrease of
polarization with increasing collision energy. Our mechanism leads to
polarization of $\bar \Lambda$ of the same sign and larger magnitude
than the polarization of $\Lambda$. The energy and charge dependence
of polarization is suggested as a sensitive probe of fine details of
QCD matter structure.

\end{abstract}


\maketitle

\section{Introduction}

The local violation \cite{Fukushima:2008xe} of discrete symmetries
in strongly interacting QCD matter is entering a new important
phase of its investigation. It started from Chiral Magnetic Effect 
(CME) \cite{Fukushima:2008xe} which uses the (C)P-violating (electro)magnetic field generated in heavy ion collisions in order to probe the (C)P-odd 
effects in QCD matter.

Current development is related to counterpart of this effect, Chiral
Vortical Effect (CVE)\cite{Kharzeev:2007tn} due to coupling to P-odd
medium vorticity leading to the induced electromagnetic and all
conserved-charge currents \cite{Rogachevsky:2010ys}, in particular
the baryonic one.

What became most important now is that P-odd effects might be observable as baryon polarization. A mechanism analogous to CVE (known as axial
vortical effect, see \cite{Kalaydzhyan:2014bfa} and references
therein) leads to an induced axial current of strange quarks which may
be converted to polarization of $\Lambda$-hyperons
\cite{Rogachevsky:2010ys,Gao:2012ix,Baznat:2013zx}. Another
mechanism of this polarization is provided  by so-called thermal
vorticity in the hydrodynamical approach \cite{Becattini:2013vja}.

Recently pioneering preliminary results on global polarization
of $\Lambda$ and $\bar \Lambda$ hyperons in peripheral $Au-Au$ collisions in
the RHIC Beam Energy Scan were released \cite{Lisa} showing a decrease
of polarization  with increasing energy in qualitative 
agreement with the prediction of ref.\cite{Rogachevsky:2010ys}. Here, we address this issue and explore the relevant
details of theoretical description. The decrease with energy is
shown to be related to the suppression of the Axial Magnetic effect
contribution in strongly correlated QCD matter found in lattice
simulations. Consequently,  accurate measurements of polarization
energy dependence may serve as a sensitive probe of strongly correlated
QCD matter.

\section{Axial anomaly and hyperon polarization}

We consider hyperon polarization as the observable related  to
vorticity and helicity \cite{Rogachevsky:2010ys}. We shall
concentrate mostly on $\Lambda (\bar \Lambda)$ hyperons, which are
produced in large numbers and their polarization may be easily
recovered from the angular distributions of their weak decays
products. These advantages are important in the current STAR
measurements \cite{Lisa}.

We explore the mechanism of generation of axial current similar to
the famous axial anomaly. In the medium described by a chemical potential
$\mu(x)$ there is a contribution to the interaction Lagrangian
\cite{Sadofyev:2010is} proportional to the appropriate conserved
charge density in the medium rest frame $\rho(x)=j_0(x)$: $$\Delta
L(x)  =\mu(x) \rho(x).$$ The Lorentz covariance allows one to
transform this expression using the  hydrodynamical four-velocity
$u_\alpha=\gamma(1, \vec v)$ where $\gamma$ is the Lorentz factor:
$$\Delta L(x)  =\mu(x) u^\alpha(x) j_\alpha(x).$$
Here, the velocity $u_\alpha(x)$ and the chemical potential $\mu(x)$
play the role of the gauge field $A(x)$ and the corresponding
coupling $g$, respectively:
\begin{equation}
g A^\beta(x) j_\beta(x) \to \mu(x) u^\alpha(x) j_\alpha(x)
\label{subst}
\end{equation}
This substitution can be applied to any diagram with the lines of
external (classical) gauge fields leading to various medium effects.
In the case of the famous anomalous triangle diagram (Fig.1) it leads to
the induced (classical) axial current.

\begin{figure}[h!]
%\includegraphics[angle=-0,width=0.4\textwidth]{vortt25_xyz_s.eps}
 %\hspace{-20mm}
\includegraphics
%[angle=-0,width=0.1\columnwidth]{helamomt.pdf}
%[width=1\textwidth]
[height=0.5\textheight] {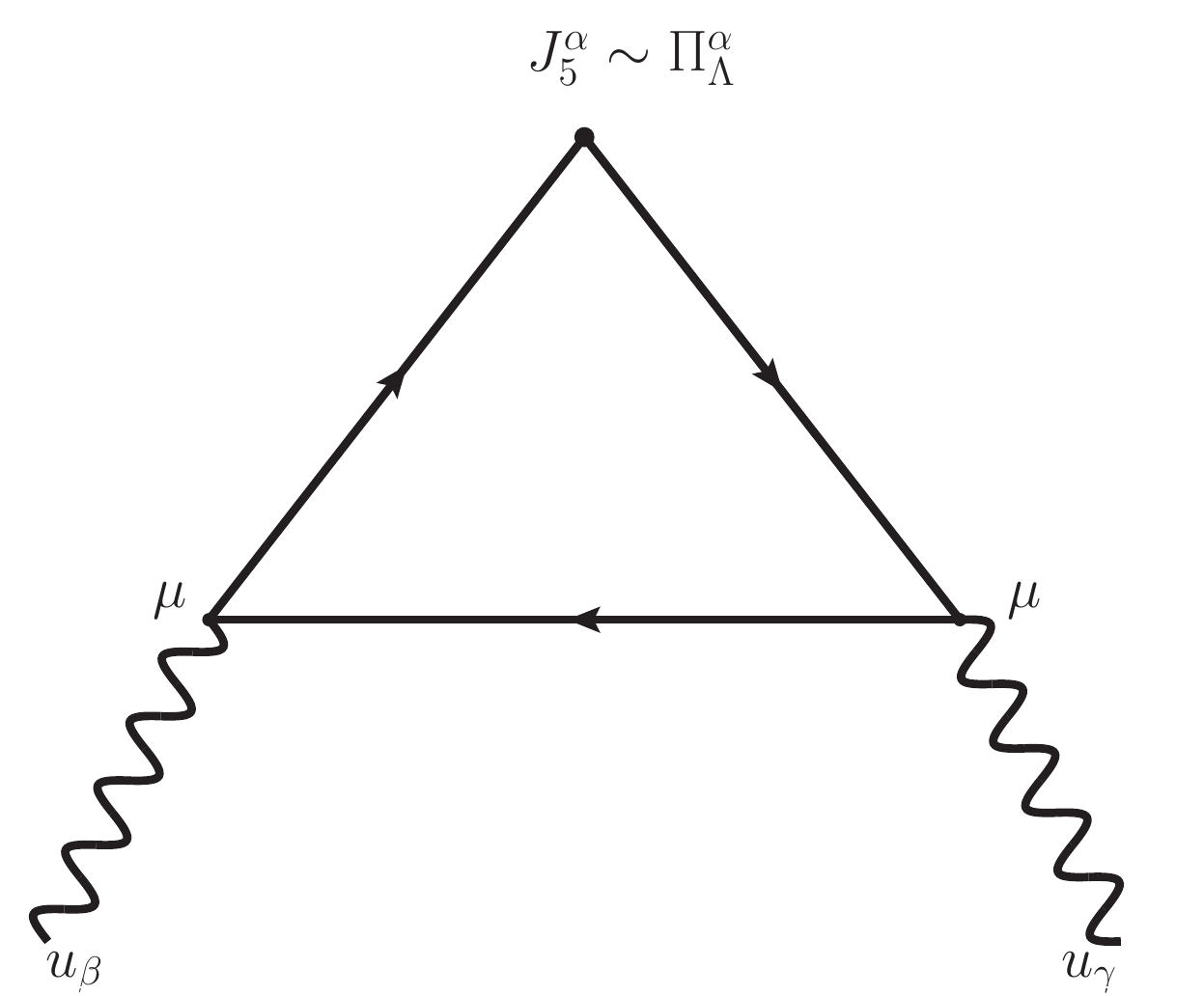} \caption{Generation of $\Lambda$
hyperon polarization via the axial anomaly}
\end{figure}

This effect is quite similar to the anomalous gluon contribution to
the nucleon spin (see e.g. \cite{Efremov:1989sn,Anselmino:1994gn} ).
The role of the gauge gluon field is played by the velocity field while the 
longitudinal polarization (helicity) of gluons corresponds, as we
will see below, to the hydrodynamic helicity.

Note that for massive quarks the anomalous contribution is partially
compensated by the "normal" one (see e.g. \cite{Efremov:1989sn}).
For a heavy quark axial current $\bar Q\gamma_\mu \gamma_5 Q$  the resulting matrix element between momentum eigenstates may be expanded to inverse
powers of the quark mass $m_Q$ \cite{Polyakov:1998rb}:
\begin{eqnarray}
\nonumber \langle p|\bar Q\gamma_\mu \gamma_5 Q|p\rangle&=&
i\frac{N_c \alpha_s }{2\pi } \varepsilon_{\mu\nu\lambda\rho} e^\nu
e^{*\rho}p^\lambda \Bigl\{ 1-\int_0^1 dx \frac{2
m_Q^2(1-x)}{m_Q^2-p^2x(1-x)}
\Bigl\} \\
&=& -i\frac{N_c\alpha_s }{12\pi } \varepsilon_{\mu\nu\lambda\rho}
e^\nu  e^{*\rho}p^\lambda\frac{p^2}{m_Q^2} + O(\frac{1}{m_Q^4})\, ,
\end{eqnarray}
where $e^\nu, \, e^{*\rho}$ are polarization vectors of these eigenstates 
and $N_c$ is
the number of colors. As far as 
the $SU(3)$ wave function of $u$ and $d$ quarks form
the spin singlet, we assume that $\Lambda$ spin is carried
predominantly by the strange quark. The strange quark may be
considered as both light (with respect to the nucleon mass) and
heavy (with respect to intrinsic higher twist scale)
\cite{Teryaev:2009zz}. When the transition to coordinate space is
performed, the Fourier transform of the corresponding matrix
elements (requiring, generally speaking, the knowledge of the anomaly
graph at arbitrary external momenta \cite{Veretin:1994dn}) should
contribute to the classical axial current. We do not expect that
these corrections could change the scale of the effect
substantially.

Let us therefore consider the classical strange axial charge
\cite{Baznat:2013zx}
\begin{eqnarray}
\label{q5s} Q_5^s=\frac{N_c}{{2 \pi^2}} \int d^3 x \,\mu_s^2(x)
\gamma^2 \epsilon^{i j k}v_{i} \partial_{j}v_ k
\end{eqnarray}
induced by the anomalous triangle diagram (Fig. 1). Here, we keep actually only the first term in
the chiral vorticity coefficient
\begin{equation}
\label{cv} c_V=\frac{\mu_s^2+\mu_A^2}{2 \pi^2}+\frac{T^2}{6},\quad
Q_5^s=N_c \int d^3 x \,c_V \gamma^2 \epsilon^{i j k}v_{i}
\partial_{j}v_ k
\end{equation}
since we assume that the chiral chemical potential $\mu_A$ is much
smaller than the strange one $\mu_s$. The temperature dependent term in eq. (\ref{cv}), related to the gravitational anomaly \cite{Landsteiner:2011iq},
can naively be considered to be quite substantial. 
However, lattice simulations \cite{Braguta:2016pwq} lead to a zero
result in the confined phase and to suppression by one order of
magnitude at high temperatures. As far as for free fermion gas the
$T^2/6$ term is recovered \cite{Buividovich:2013jba} for large
lattice volume at fixed temperature, the above-mentioned suppression
should be attributed to the correlation effects.

In order to relate the strange axial charge $Q_5^s$ (\ref{q5s}) to
hydrodynamical quantities one can use the mean-value theorem to
evaluate it \cite{Baznat:2013zx}:
\begin{eqnarray}
\label{mean1} Q_5^s=\frac{<\mu^2 \gamma^2> N_c H}{2 \pi^2},
\end{eqnarray}
where hydrodynamical helicity
$$H \equiv \int d^3 x  (\vec v \cdot \vec w)$$
is the integrated projection of the velocity $\vec v$ to the
vorticity $\vec w= curl \vec v$.

Note that the hydrodynamic helicity is related to the zeroth component
($\mu=0$) of the four-current
\begin{equation}
K^\mu (x) =\epsilon^{\mu \nu \rho \gamma }u_{\nu} (x)
\partial_{\rho} u_\gamma (x),\,\ H=\int d^3 x
\frac{K^0(x)}{\gamma^2}. \label{K}
\end{equation}
Coming back to the similarity with the spin crisis, this is the analog of
the topological current, related \cite{Efremov:1989sn} (in the axial
gauge) to the gluon polarization.

The space components of (\ref{K}) for slow enough fields in the
non-relativistic approximation are related to the vorticity vector:
$$K^i (x)|_{\gamma \to 1} =  2 \epsilon^{i j k} \partial_{j} v_k(x)= 2 \omega^i,$$
while the exact relation involves relativistic corrections. To avoid such complications, here we discuss the complementary approach \cite{Baznat:2015eca} 
relying on the $K^0 (x)$, helicity and axial charge. To pass from the classical 
charge (we are always dealing with) to the quantum matrix elements 
we will use the analogy with the conserved current case when the 
conserved charge $Q$ appears in the symmetric one-particle matrix 
elements of the current
\begin{equation}
\label{clq} <p_n|j^0(0)|p_n>=2 p^0_n Q_n.
\end{equation}
To calculate the (quantum) average charge per particle resulting
from the latter equation it is sufficient to divide the total {\it
classical} charge by particles' number $N$:
\begin{equation}
\label{clq1} <Q> \equiv \frac{\sum_{n=1}^N Q_n}{N}=\frac{\int d^3 x
\, j^0_{class}(x)}{N}.
\end{equation}
Passing now to the axial charge case, let us note that the symmetric
matrix element of quark axial current of the flavor $i$ is related
to the fraction $a_i$ of a hadron covariant polarization $\Pi^\mu \,
(p_\mu \Pi^\mu=0)$ carried by that quark:
$$<p_n,\Pi_n|j_{5,i}^0(0)|p_n,\Pi_n>=2 a_{i,n} m_n \Pi_n^0,$$
where $m_i$ is a hadron $i$ mass. By analogy with
(\ref{clq}--\ref{clq1}), the axial charge should correspond to
$$Q_{5,i,n} \to \frac{m_n a_{i,n} \Pi_n^0}{p_n^0}.$$
The average polarization is then
 $$< \frac{a_i m \Pi^0}{p_0}>\equiv \frac{\sum_{n=1}^N  \frac{m_n a_{i,n}\Pi^0_n}{p^0_n}}{N}= \frac{\int d^3 x \, j^0_{5,class}(x)}{N}.$$

In what follows we will consider all $N$ particles to be $\Lambda
\,(\bar \Lambda)$ and put $a_i m/p_0=1$ assuming (following the
above-mentioned $SU(3)$ based arguments) that strange (anti)quarks
carry all the polarization and considering the $\Lambda$ mass is large
enough with respect to its momentum. The latter assumption is a reasonable
approximation at the NICA energies ($\sqrt{s_{NN}}=4 - 11 GeV$) and
will provide a lower bound estimate of the polarization.

The helicity shows the phenomenon of separation
\cite{Baznat:2013zx,Teryaev:2014qia} so that its sign is changed at
the two sides of the reaction plane. As a result, the axial charge and
the zeroth component of the hyperon polarization also manifest such a
phenomenon of sign change.

%Selecting the axial charge
%related to the particles in the definite rapidity %or transverse
%momentum interval, the respective dependence of %polarization may be
%also obtained.

As the axial charge is related to the zeroth component of the hyperon
polarization in the laboratory frame $\Pi_0^{lab}$, the transformation
to hyperon rest frame must be performed \cite{Baznat:2015eca}.
Taking into account that the polarization pseudovector should be
directed along the $y-$axis transverse to the reaction plane (as it
has to be collinear to the angular momentum pseudovector), one gets
for the components of the laboratory frame polarization
\begin{equation}
\label{Piv} \Pi^{\Lambda,lab} = \big(\Pi_0^{\Lambda,lab},
\Pi_x^{\Lambda,lab}, \Pi_y^{\Lambda,lab}, \Pi_z^{\Lambda,lab} \big)
= \frac{\Pi_0^{\Lambda}}{m_{\Lambda}}\big( p_y, 0, p_0, 0 \big).
\end{equation}
One can use both its time $\Pi_0^{\Lambda,lab}$ and space
$\Pi_y^{\Lambda,lab}$ components to recover the polarization in the
hyperon rest frame. Note that for the time component the factor $p_y$
changes sign at two sides of reaction plane compensating the
sign change due to helicity separation. At the same time, that
factor is absent for space component. As we discussed above, this
component is related to the vorticity rather than helicity and
should not show the sign change. Therefore, the use of either time
or space component of $\Pi_0^{\Lambda,lab}$ should lead to equal
polarization at both sides of the reaction plane. Leaving the
exploration of space component for future work we use the time
component and get:
\begin{equation}
\label{ratio} <\Pi_0^{\Lambda}>\, = \, \frac{m_\Lambda \,
\Pi_0^{\Lambda,lab}}{p_y}\,= \, <\frac{m_{\Lambda}}{N_{\Lambda}
\,p_y }> Q_5^s \,\equiv \, <\frac{m_{\Lambda}}{N_{\Lambda} \,p_y }>
\frac{N_c}{{2 \pi^2}} \int d^3 x \,\mu_s^2(x) \gamma^2 \epsilon^{i j
k}v_{i} \partial_{j}v_ k.
\end{equation}

The appearance of $p_y$ in the denominator is not dangerous, as the
particles  with zero transverse momentum do not have also the time
component of polarization.

%The extra factor $p_y$ provides the extra source of the sign change
%at the opposite sides of the reaction plane compensating the sign
%change of the helicity and axial charge, so that the resulting sign
%of the polarization remains the same.

The average polarization was firstly roughly  estimated just by
dividing $Q_5^s$ (\ref{mean1}) by the number of $\Lambda$
leading \cite{Baznat:2013zx} to a value of about
$1\%$, later confirmed \cite{Baznat:2015eca} by more detailed
simulations, which are compatible with the current STAR data
\cite{Lisa}.

The appearance of $\mu^2$ in eqs. (\ref{q5s}) and
(\ref{ratio}), related to the positive C-parity of axial current,
immediately leads to the same expressions for axial charge of
strange quarks and antiquarks. As far as there is a smaller number
of $\bar \Lambda$s than of $\Lambda$s, so that the same
axial charge should be distributed among smaller number of
antiquarks comparing with the number of quarks, the corresponding
factor in the denominator in eq. (\ref{ratio}) is smaller for $\bar
\Lambda$s, which results in an increase of the effect for the latter.
Thus, one could expect that the polarization of $\bar \Lambda$ has
to be of the same sign but of a larger magnitude than the polarization
of $\Lambda$, which is compatible with the quite recent STAR data
\cite{Lisa}. This effect might be partly compensated by the fact,
that a larger amount of axial charge in the case of strange
antiquarks might be carried by more numerous $K^*$-mesons.

As far as the strange chemical potential is rapidly decreasing with
energy, this provides a natural explanation of the observed hint
\cite{Lisa} for decrease of polarization with energy. More accurate
measurements of $\Lambda$ and $\bar \Lambda$ polarization at RHIC,
and at NICA and FAIR in future, might allow one to test the
suppression of $T^2$ term and, at best, even to check experimentally
the magnitude of its theoretically predicted coefficient and search
for possible extra \cite{Becattini:2013vja} contributions.

%\vspace*{-2cm}
\section{Conclusions and Outlook}

%\vspace*{-1cm}
The generation of polarization by the anomalous mechanism (Axial
Vortical Effect) naturally explains several features of the observed
data.

\begin{itemize}

\item {The decrease of chemical potential with energy leads to the
decrease of polarization. An additional source of decrease is
provided by the energy dependence of the helicity which was earlier
found \cite{Baznat:2013zx} to be maximal in the NICA energy range.
 The contribution related to the gravitational
anomaly proportional to $T^2$ may be suppressed in strongly
correlated matter. Moreover, the accurate measurements of the energy
dependence of polarization should allow to separate the
gravitational anomaly contribution and test the degree of its
suppression in strongly correlated QCD matter.}

\item The proportionality of the polarization to the square of the chemical
potential related to C-even parity of axial current leads to the
same sign of polarization of $\Lambda$ and $\bar \Lambda$ hyperons.
The smaller number of the latter should result in a larger fraction
of the axial charge corresponding to each anti-hyperon and to a larger
absolute value of polarization. Detailed numerical simulations
may allow to quantify this prediction. Accurate measurements of
$\Lambda$ and $\bar \Lambda$ polarization should allow to check
these predictions and provide an additional check of the gravitational
anomaly related contribution.

\end{itemize}

%{}~

{\bf Acknowledgments} We are indebted to M. Baznat, K. Gudima, O.
Rogachevsky, and R. Usubov for long-lasting collaboration in the
investigations of P-odd effects in heavy-ion collisions. We are
grateful to F. Becattini, T. Bir\'o, V. Braguta, P. Buividovich,
A.~Efremov, V. Goy, V. Kekelidze, M. Lisa, I. Tserruya and X.N. Wang for
stimulating discussions and useful comments.

%{%}~

This work was supported in part by the Russian Foundation for Basic
Research, Grant No. 14-01-00647.

%\begin{thebibliography}{}

\end{document}